\documentclass[12pt]{iopart}
\usepackage{graphicx,epsf}
\usepackage{dcolumn}
\usepackage{bm}

\begin{document}
\title{Neutrino Masses from Cosmological Probes}
\author{\O ystein Elgar\o y\dag\  and Ofer Lahav \ddag}
\address{\dag\ Institute of theoretical astrophysics, University of Oslo, P.O. Box 1029, N-0315 Oslo, Norway}
\address{\ddag\ Department of Physics and Astronomy, University College London, Gower Street, London WC1E 6BT, UK}
\eads{\mailto{oelgaroy@astro.uio.no};  \mailto{lahav@star.ucl.ac.uk}}

\begin{abstract}

There is a renewed interest in constraining the sum of the masses of
the three neutrino flavours  by using cosmological measurements.  
Solar, atmospheric, 
reactor and accelerator neutrino experiments have confirmed 
neutrino oscillations,
implying that neutrinos have non-zero mass, but without pinning down
their absolute masses.  While it is established that the effect of
light neutrinos on the evolution of cosmic structure is small, the upper
limits derived from large-scale structure could help significantly to
constrain the absolute scale of the neutrino masses.  It is also
important to know the sum of neutrino masses as it is degenerate with
the values of other cosmological parameters, e.g. the amplitude of
fluctuations and the primordial spectral index.  
A summary of
cosmological neutrino mass limits is given.  
Current results from
cosmology set an upper limit on the sum of the neutrino masses of
$\sim 1\;{\rm eV}$, somewhat depending on the data sets used in the
analyses and assumed priors on cosmological parameters.  
It is important to
emphasize that the total neutrino mass (`hot dark matter') is derived
assuming that the other components in the universe are baryons, cold
dark matter and dark energy.  We assess the impact of
neutrino masses on the matter power spectrum, the cosmic microwave
background, peculiar velocities and gravitational lensing.  We also
discuss future methods to improve the mass upper limits by an order of
magnitude.

\end{abstract}

\maketitle

\section{Introduction}

The connection between neutrino masses and cosmic structure formation 
was realised in the 1970's by Zeldovich and others, 
but for a long time cosmologists were  
mostly interested in neutrino masses in the $\sim 10\;{\rm eV}$ range, 
i.e. massive enough to make up all of the dark matter.  
The downfall of the top-down `hot dark matter' 
scenario of structure formation, 
and the fact that no evidence for neutrino masses existed before 
Super-Kamiokande detected oscillations of atmospheric neutrinos in 1998, 
explains why there was very little continuous interest 
in this sub-field.  However, the detection of neutrino oscillations 
showed that neutrinos indeed have a mass, i.e. 'hot dark matter' does exist,
even if only in a very small amount.

The wealth of new data from  the cosmic microwave background 
(CMB) and large-scale structure (LSS) in the last few years  
indicate that we live in a flat Universe where  
$\sim 70\;\%$ of the mass-energy density is in the form of dark energy, 
with matter making up the remaining 30 \% .   
The WMAP data combined with 
other large-scale structure data \cite{wmap,sdsspap1} give impressive 
support to this picture. 
Furthermore, the baryons contribute only a fraction $f_{\rm b} 
= \Omega_{\rm b}/\Omega_{\rm m} \sim 0.15$ ($\Omega_{\rm b}$ and 
$\Omega_{\rm m}$ are, respectively, the contribution of baryons and 
of all matter to the total density in units of the critical density 
$\rho_{\rm c} =  3H_0^2 / 8\pi G = 1.879\times 10^{-29}h^2\;{\rm g}\,{\rm cm}
^{-3}$, where $H_0 = 100 h \;{\rm km}\,{\rm s}^{-1}\,{\rm Mpc}^{-1}$ is 
the present value of the Hubble parameter) of this, so that 
most of the matter is dark. 
The exact nature of the dark matter in the Universe is still 
unknown.  Relic neutrinos are abundant in the 
Universe, and from the observations of oscillations of 
solar and atmospheric neutrinos, as well as in the reactor-based KamLAND experiment and in the accelerator-based long-baseline K2K experiment, we know that neutrinos have 
a mass \cite{sage,sno1,sno2,macro,gno,homestake,superk,gallex,kamland1,k2k}
and will make 
up a fraction of the dark matter.   The KamLAND collaboration has also 
reported evidence for spectral distortion in the $\overline{\nu}_{e}$ 
spectrum \cite{kamland2}, 
further strengthening the case for neutrino oscillations and 
allowing rather strict bounds on the the neutrino oscillation parameters 
to be found.  
However, the oscillation experiments 
can only measure differences in the squared masses of the neutrinos, 
and not the absolute mass scale, so they cannot, at least not without 
extra assumptions,  tell us 
how much of the dark matter is in neutrinos.  
From general arguments on structure formation in the Universe we know 
that most of the dark matter has to be cold, i.e. non-relativistic 
when it decoupled 
from the thermal background.  Neutrinos with masses on the 
eV scale or below will be a hot component of the dark matter.  
If they were the dominant dark-matter component, structure in the 
Universe would have formed first at large scales, and smaller structures 
would form by fragmentation (the `top-down' scenario).  
However, the combined observational 
and theoretical knowledge about large-scale structure gives strong evidence 
for the `bottom-up' picture of structure formation, i.e. structure formed 
first at small scales.  Hence, neutrinos cannot make up 
all of the dark matter (see e.g. \cite{primack} for a review).       
Neutrino experiments give some constraints on how much of the dark 
matter can be in the form of neutrinos.
Studies of the energy spectrum in 
tritium decay \cite{mainz} provide an upper limit on the effective 
electron neutrino mass involved in this process of 
2.2 eV (95 \% confidence limit).
For the effective neutrino mass scale involved in neutrino less 
double beta decay a range 0.1-0.9 eV has been inferred from the 
claimed detection of this process \cite{klapdor1,klapdor2}.  
If confirmed, this result would not only show that neutrinos are   
Majorana particles (i.e. their own antiparticles), but also that 
the neutrino masses are in a range where they are potentially 
detectable with cosmological probes.    

The structure of this review is as follows.  
Sections 2 and 3 describe the 
role of massive neutrinos in structure formation 
and for the CMB anisotropies, respectively.
In section 4 we give an overview of 
recent cosmological neutrino mass limits.  Section 5 discusses 
other methods for 
constraining neutrino masses.  
We discuss challenges for the future in section 6.

\section{Massive neutrinos and structure formation}

The relic abundance of neutrinos in the Universe today is 
straightforwardly found from the fact that they continue to 
follow the Fermi-Dirac distribution after freeze-out, and their 
temperature is related to the CMB temperature $T_{\rm CMB}$ today by  
$T_\nu = (4/11)^{1/3}T_{\rm CMB}$, giving 
\begin{equation}
n_\nu = \frac{6\zeta(3)}{11\pi^2}T_{\rm CMB}^3, 
\label{eq:relicabund}
\end{equation}
where $\zeta(3)\approx 1.202$, 
which gives $n_\nu \approx 112\;{\rm cm}^{-3}$ at present. 
By now, massive neutrinos will have become non-relativistic, so that 
their present contribution to the mass density can  
be found by multiplying $n_\nu$ with the total mass 
of the neutrinos $m_{\nu,\rm tot} \equiv \sum m_{\nu}$, giving 
\begin{equation}
\Omega_\nu h^2 = \frac{m_{\nu,\rm tot}}{94\;{\rm eV}},
\label{eq:omeganu}
\end{equation}
for $T_{\rm CMB}=2.726\;{\rm K}$.  Several effects could  
modify this simple relation.  If any of the neutrino chemical 
potentials were initially non-zero, or there were a sizable  
neutrino-antineutrino asymmetry, this would increase 
the energy density in neutrinos and give an additional contribution 
to the relativistic  energy density.  However, from Big Bang Nucleosynthesis 
(BBN) one gets a very tight limit on the electron neutrino chemical potential, 
since the electron neutrino is directly involved in the processes that 
set the neutron-to-proton ratio.  Also, within the standard three-neutrino 
framework one can extend this limit to the other flavours as well.   
The  results of the KamLAND experiment \cite{eguchi} 
confirmed the Large Mixing Angle (LAM) solution for the solar 
neutrino oscillations, and combined with the atmospheric data indicating 
maximal mixing in this sector, it has been shown that flavour equilibrium 
is established between all three neutrino species before the epoch 
of BBN \cite{lunardini,dolgov,ywong,kevork}, so that the BBN constraint on the 
electron neutrino asymmetry 
applies to all flavours, which in turn implies that the lepton 
asymmetry cannot be large enough to give a significant contribution 
to the relativistic energy density.  One should note, however, that 
these bounds can be evaded if there is an additional source of energy density 
to compensate the $\nu_e$ asymmetry \cite{kneller}.  
Analyses of WMAP and 2dFGRS 
data give independent, although not quite as strong, evidence for 
small lepton asymmetries \cite{steen3,elena}.
Within the standard picture, 
equation (\ref{eq:relicabund}) should be accurate, and therefore 
any constraint on the cosmic mass density of neutrinos should translate 
straightforwardly into a constraint on the total neutrino mass, 
according to equation (\ref{eq:omeganu}).  
If a fourth, light `sterile' neutrino exists, sterile-active oscillations 
would modify this conclusion, and could also have implications for 
e.g. BBN \cite{kevork2}.  See \cite{cirelli} for a review.  
No sterile neutrinos are required to 
explain the solar and atmospheric neutrino oscillation data 
\cite{pakvasa}, and the only hint so far comes from the 
possible detection of $\overline{\nu}_\mu \rightarrow \overline{\nu}_e$ 
oscillations with a small mixing angle and a mass-square difference 
$\sim 1\;{\rm eV}^2$ at the Liquid Scintillating Neutrino Detector (LSND)  
\cite{LSND}.   Since there are only two independent mass-squared 
differences in the standard three-neutrino scenario, and they are 
orders of magnitude smaller, this hints at the existence of a 
fourth, light sterile neutrino.  However, this has found to be highly 
disfavoured by the solar and atmospheric neutrino oscillation data 
\cite{maltoni1,maltoni2}.  The status 
of the LSND results will in the near future be clarified  by  
the MiniBooNE experiment \cite{mboone}.

In a recent paper Beacom, Bell \& Dodson \cite{noneut} pointed out that if neutrinos
couple to a light scalar field, they might annihilate as they become
non-relativistic.  If this were the case, massive neutrinos would have
a negligible effect on the matter power spectrum.  However, Hannestad
\cite{hannestad0411475}
has pointed out that such a scenario would have
a marked effect on the CMB power spectrum and is strongly disfavoured
by the current data.   We will therefore assume in the following that
no such non-standard couplings of neutrinos exist.

\begin{figure}
\begin{center}
\includegraphics[width=84mm]{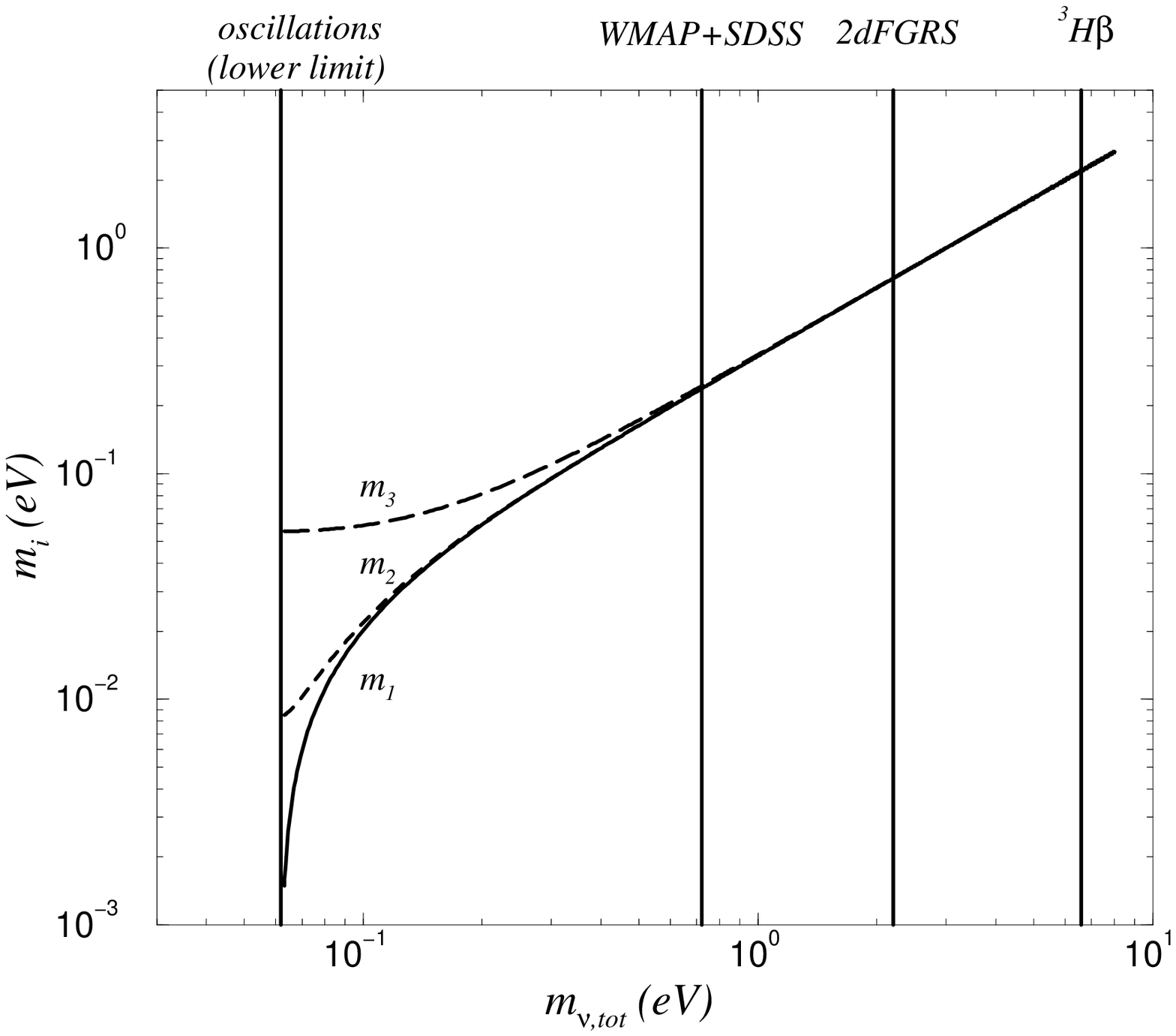}
\includegraphics[width=84mm]{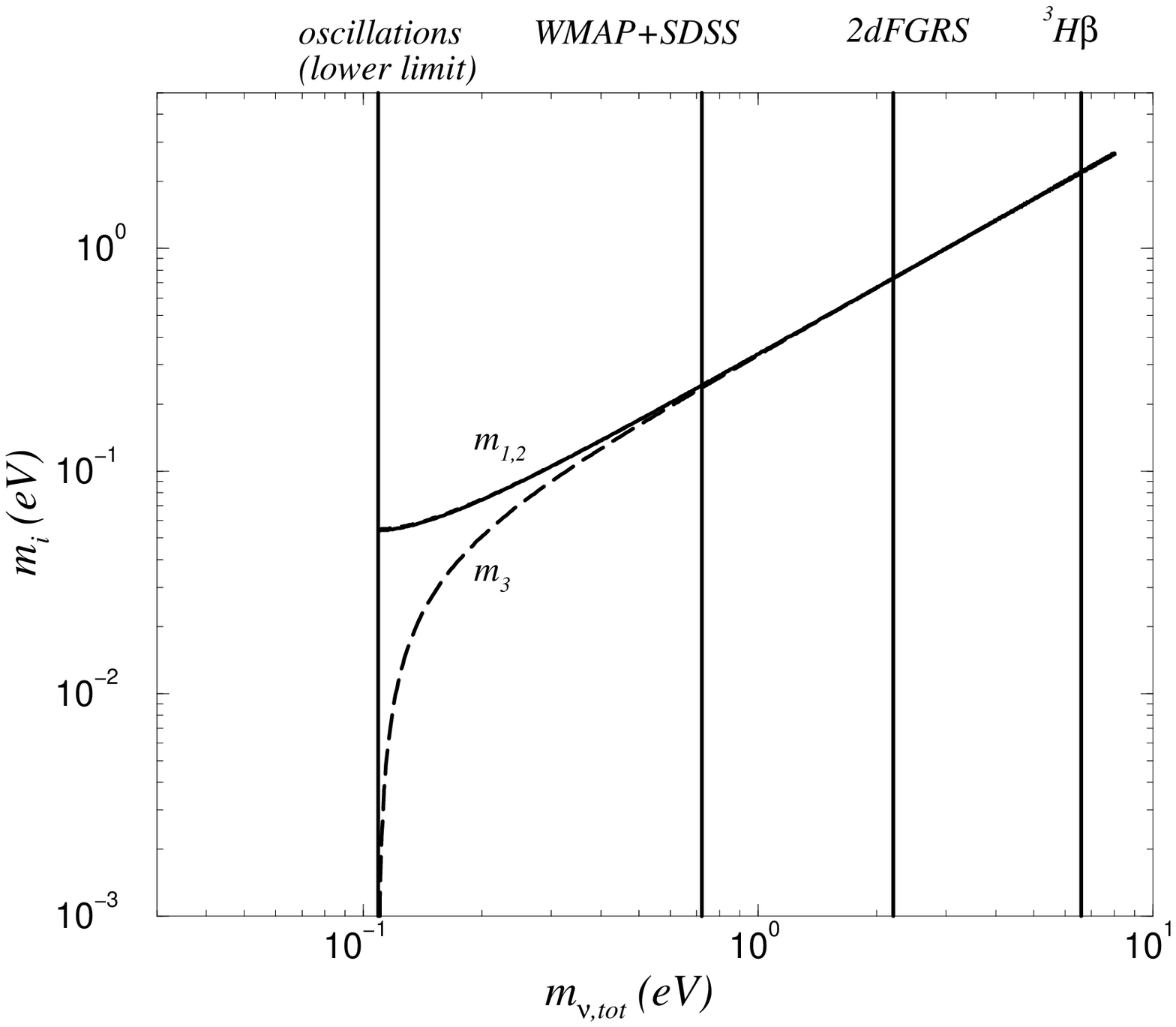}
\end{center}
\caption{Neutrino mass eigenvalues as functions of $m_{\nu,\rm tot}$ for 
the cases of normal (top panel) and inverted (bottom panel) hierarchies.  
The vertical line marked `oscillations' is the lower limit derived from 
the measured mass-squared differences. 
The vertical line marked `WMAP+SDSS' is the recent limit derived 
in \cite{seljakbias} from WMAP+SDSS, the vertical line marked 
'2dFGRS' is the limit from the 2dFGRS derived in \cite{elgar}, and 
the line marked 
`$^3{\rm H}\beta$' is the upper limit from Tritium $\beta$ decay.}
\label{fig:fig1}
\end{figure}

Finally, we assume that the neutrinos are 
nearly degenerate in mass.  Current cosmological observations 
are sensitive to neutrino masses  $\sim 1\;{\rm eV}$ or greater.   
Since the mass-square differences are
small, the assumption of a degenerate mass hierarchy is therefore 
justified.  This is illustrated in figure \ref{fig:fig1}, 
where we have plotted the mass eigenvalues $m_1,m_2,m_3$ as 
functions of $m_{\nu,\rm tot} = m_1 + m_2 + m_3$ for 
$\Delta m_{21}^2 = 7\times 10^{-5}\;{\rm eV}^2$ (solar) and 
$\Delta m_{32}^2 = 3 \times 10^{-3}\;{\rm eV}^2$ (atmospheric), 
for the cases of a normal hierarchy ($m_1 < m_2 < m_3$), and 
an inverted hierarchy ($m_3 < m_1 < m_2$).  As seen in the figure, 
for $m_{\nu,\rm tot} > 0.4\;{\rm eV}$ the mass eigenvalues are 
essentially degenerate. 

Here we look at cosmological models 
with four components: baryons, cold dark matter, massive neutrinos, 
and a cosmological constant.  Furthermore, we restrict ourselves 
to adiabatic, linear perturbations.  The basic physics is then 
fairly simple.  A perturbation mode of a given wavelength $\lambda$ 
can grow if it is greater than the Jeans wavelength $\lambda_{\rm J}$ 
determined by the balance of gravitation and pressure, or rms velocity 
in the case of massless particles.  Above the 
Jeans scale, perturbations grow at the same rate independently of 
the scale.  Long after 
matter-radiation equality, all interesting scales are above 
$\lambda_{\rm J}$ and grow at the same rate, and in models where 
all the dark matter is cold, the  time and scale dependence of the 
power spectrum can therefore be 
separated at low redshifts.  Light, massive neutrinos can, however, move 
unhindered out of regions below a certain limiting length scale, 
and will therefore tend to damp a 
density perturbation at a rate 
which depends on their rms velocity.    
The presence of massive neutrinos 
therefore introduces a new length scale, given by the size of the 
co-moving Jeans length when the neutrinos became non-relativistic.  
In terms of the comoving wavenumber, this is given by  
\begin{equation}
k_{\rm nr} = 0.026\left(\frac{m_{\nu}}{1\;{\rm eV}}\right)^{1/2}
\Omega_{\rm m}^{1/2}\;h\,{\rm Mpc}^{-1},
\label{eq:knr}
\end{equation}
for three equal-mass neutrinos, each with mass $m_\nu$.  
The growth of Fourier modes with 
$k > k_{\rm nr}$ will be suppressed because of neutrino free-streaming.  
The free-streaming scale varies with the cosmological epoch,
and the scale and time dependence of the power spectrum cannot 
be separated, in contrast to the situation for models with cold dark 
matter only.

The transfer functions of the perturbations in the various components 
provide a convenient way of describing their evolution on different 
scales.  Using the redshift $z$ to measure time, the transfer function 
is formally defined as   
\begin{equation}
T(k,z) = \frac{\delta(k,z)}{\delta(k,z=z_*)D(z_*)}
\label{eq:tfunc}
\end{equation}
where $\delta(k,z)$ is the density perturbation with wavenumber $k$ 
at redshift $z$, and $D$ is the linear growth factor.   
The normalization redshift $z_*$ corresponds to a time long before 
the scales of interested have entered the horizon.  
The transfer function thus gives the amplitude of 
a given mode $k$ at redshift $z$ relative to its initial value, and 
is normalized so that $T(k=0,z)=1$.  The power spectrum of the matter 
fluctuations can be written as 
\begin{equation}
P_{\rm m}(k,z)= P_*(k)T^2(k,z),
\label{eq:pmatter}
\end{equation}
where $P_*(k)$ is the primordial spectrum of matter fluctuations, 
commonly assumed to be a simple power law $P_*(k)=Ak^{n}$, where 
$A$ is the amplitude and the spectral index $n$ is close to 1.  
It is also common to define power spectra for 
each component, see \cite{eisenstein} for a discussion.  
Note that the transfer functions and power spectra are insensitive 
to the value of the cosmological constant as long as it does not 
shift the epoch of matter-radiation equality significantly.  

\begin{figure}
\begin{center}
\includegraphics[width=84mm]{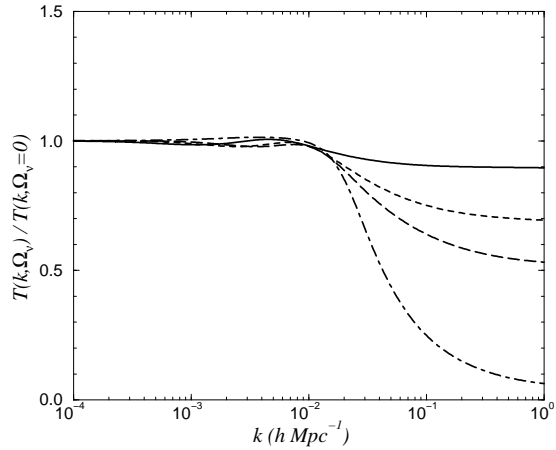}
\end{center}
\caption{Ratio of the transfer functions (at $z=0$) for various values 
of $\Omega_\nu$ to the one for $\Omega_\nu = 0$.  The other parameters 
are fixed at $\Omega_{\rm m}=0.3$, $\Omega_{\rm b}=0.04$, $h=0.7$.  
The solid line is for neutrino mass per flavour $m_\nu = 0.1\;{\rm eV}$, 
the dashed line is for 
$m_\nu=0.3\;{\rm eV}$, the long-dashed line is for $m_\nu=0.5\;{\rm eV}$, 
and the dot-dashed line corresponds to $m_\nu=2\;{\rm eV}$.}
\label{fig:transfer}
\end{figure}

The transfer function is found by solving   
the coupled fluid and Boltzmann equations for the various components.  
This can be done using one of the publicly available codes, e.g. 
CMBFAST \cite{zaldarriaga} or CAMB \cite{camb}.  
In figure \ref{fig:transfer} we show the transfer functions for models 
with $\Omega_{\rm m}=0.3$, $\Omega_{\rm b}=0.04$, 
$h=0.7$ held constant, but with varying neutrino mass per flavour 
$m_\nu \equiv m_{\nu,{\rm tot}}/3$.  One can 
clearly see that the small-scale suppression of power becomes more 
pronounced as the neutrino fraction $f_{\nu}\equiv \Omega_{\nu}/\Omega_
{\rm m}$ increases.  

\begin{figure}
\begin{center}
\includegraphics[width=84mm]{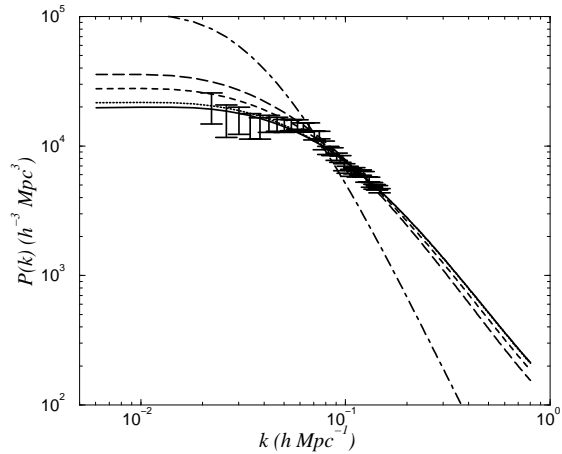}
\end{center}
\caption{Power spectra for neutrino mass per flavour $m_\nu = 0$ (full line), 
$m_\nu=0.1$ (dotted line), $m_\nu=0.3$ (dashed line), 
$m_\nu = 0.5$ (long-dashed line), and $m_\nu=3\;{\rm eV}$ (dot-dashed line).   
The other parameters 
are fixed at $\Omega_{\rm m}=0.3$, $\Omega_{\rm b}=0.04$, $h=0.7$. 
The vertical bars are the 2dFGRS power spectrum data points.}  
\label{fig:fig3}
\end{figure}

The effect is also seen in the power spectrum, as shown in 
figure \ref{fig:fig3}.  Note that the power spectra shown in 
the figure have been convolved with the 2dFGRS window function, 
as described in \cite{percival}.  
Furthermore, we have taken the possible bias of the distribution 
of galaxies with respect to that of the dark matter into 
account by leaving 
the overall amplitude of each  power spectrum as a free 
parameter to be fitted to the 2dFGRS power spectrum data (the 
vertical bars in the figure).  For a discussion of bias in 
the context of neutrino mass limits, see \cite{olahav}.    
Because the errors on the data points are smaller at small scales, 
these points are given most weight in the fitting, and hence the power spectra 
in the figure actually deviate more and more from each other on {\it large} 
scales as $m_\nu$ increases.  One can see from the figure that a 
neutrino mass of $m_\nu = 0.5\;{\rm eV}$ or larger is in conflict with 
the data.  
The suppression of the power spectrum on small 
scales is roughly proportional to $f_{\nu}$: 
\begin{equation}
\frac{\Delta P_{\rm m}(k)}{P_{\rm m}(k)} \approx -8f_\nu .
\label{eq:fnusuppr}
\end{equation}  
This result can be understood qualitatively 
from the equation of linear growth of 
density perturbations and the fact that only a fraction $(1-f_\nu)$ 
of the matter can cluster when massive neutrinos are present \cite{bond}.

\section{Constraints from the CMB alone}

\begin{figure}
\begin{center}
\includegraphics[width=84mm]{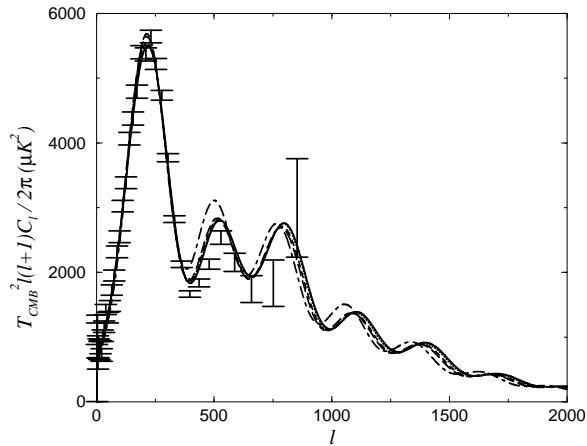}
\end{center}
\caption{CMB power spectra for neutrino mass per flavour 
$m_\nu = 0$ (full line), 
$m_\nu=0.1$ (dotted line), $m_\nu=0.3$ (dashed line), 
$m_\nu = 0.5$ (long-dashed line), and $m_\nu=3\;{\rm eV}$ (dot-dashed line).   
The other parameters 
are fixed at $\Omega_{\rm m}=0.3$, $\Omega_{\rm b}=0.04$, $h=0.7$.  
The vertical bars are the WMAP power spectrum data points}  
\label{fig:fig4}
\end{figure}

\begin{figure}
\begin{center}
\includegraphics[width=84mm]{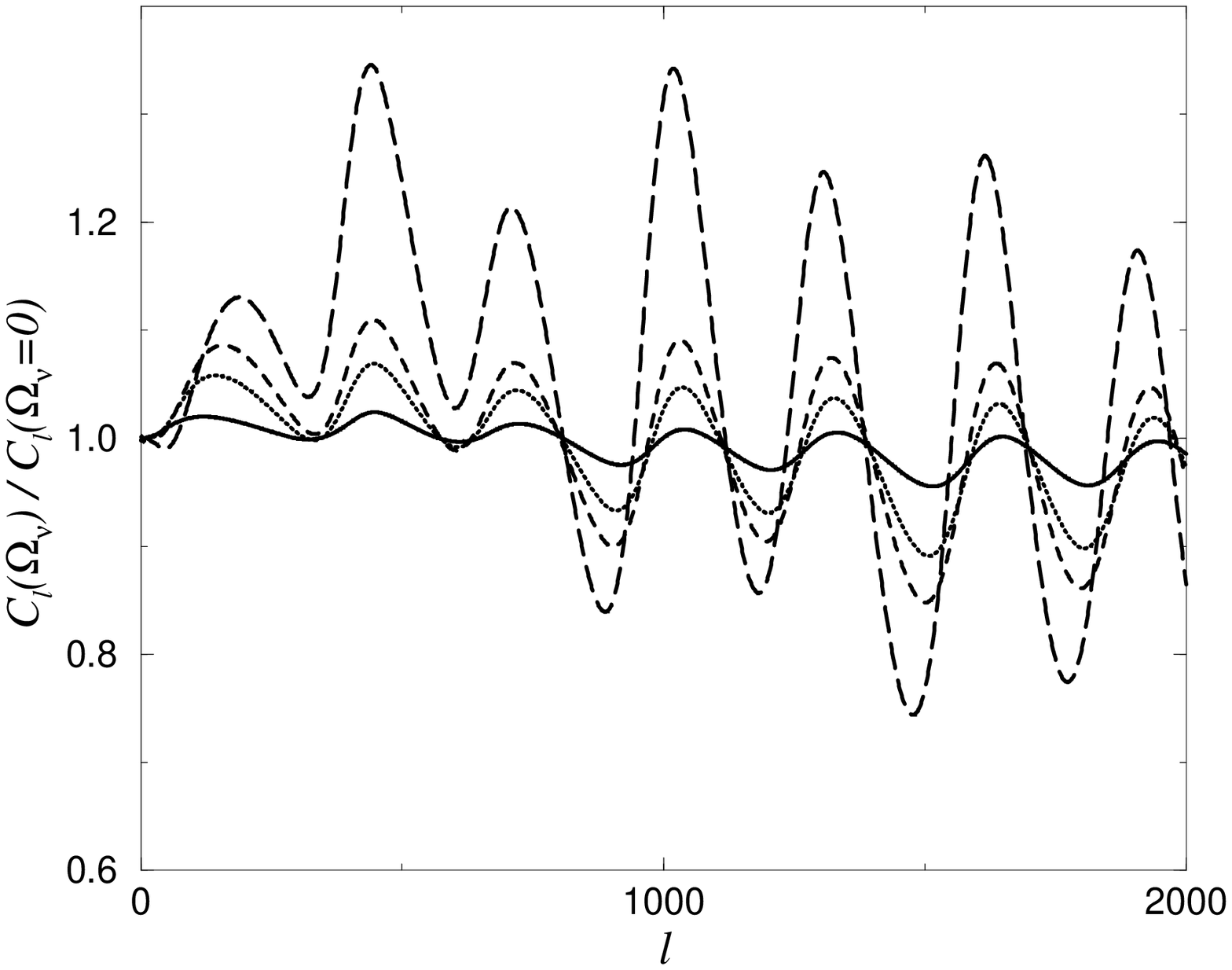}
\includegraphics[width=84mm]{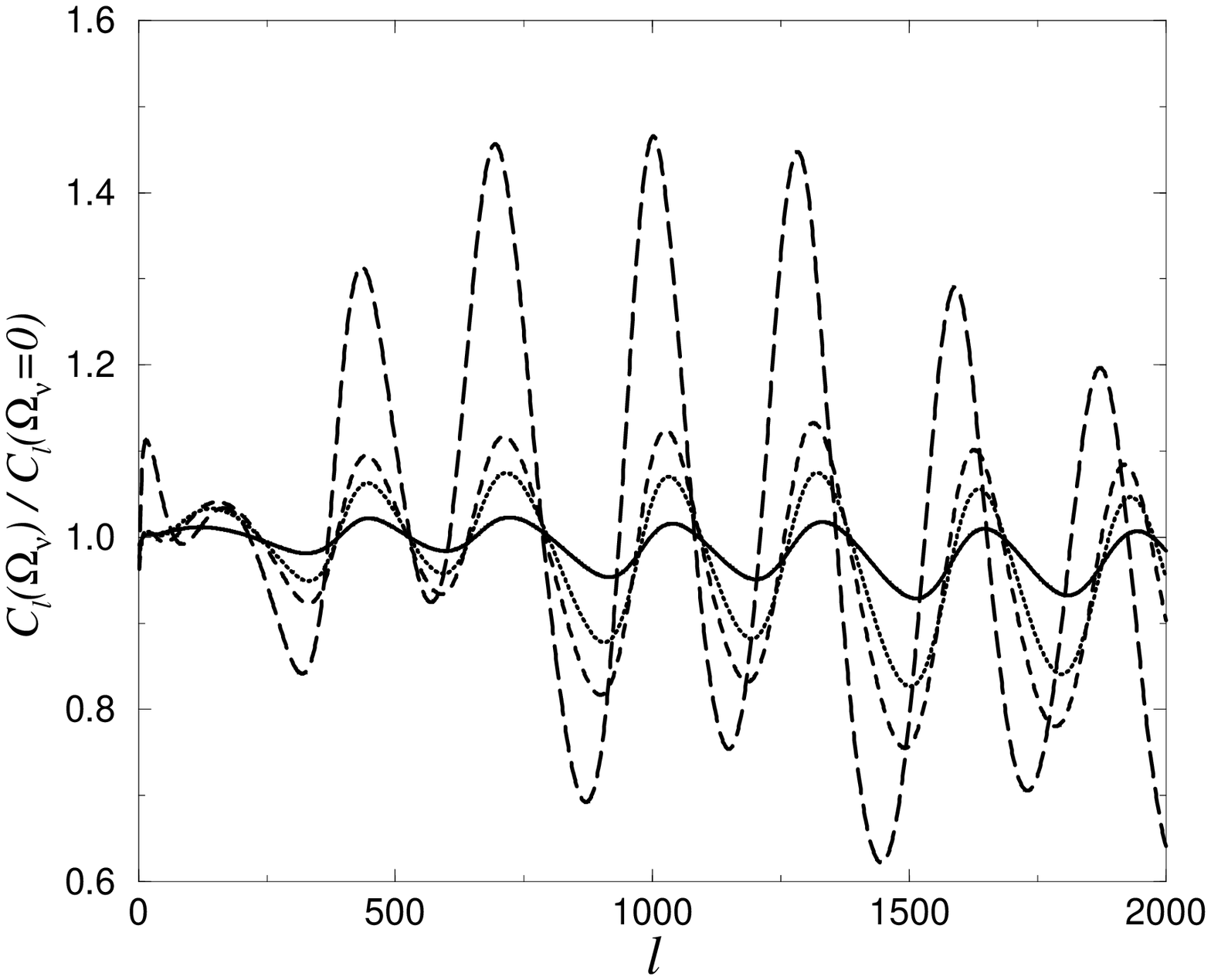}
\end{center}
\caption{
{\it Top} : The ratio of CMB angular power spectra for various values 
of $\Omega_\nu$ to the one for $\Omega_\nu = 0$.  The other parameters 
are fixed at $\Omega_{\rm m}= 
\Omega_{\rm b} + \Omega_{\rm c} + \Omega_{\nu}  =0.3$, 
$\Omega_{\rm b}=0.04$, $\Omega_{\Lambda} = 0.7$ $h=0.7$.  
The full line is for neutrino mass per flavour $m_\nu = 0.1\;{\rm eV}$, the dotted line is for 
$m_\nu=0.3\;{\rm eV}$, the dashed line is for $m_\nu=0.5\;{\rm eV}$, 
and the long-dashed line corresponds to $m_\nu=2\;{\rm eV}$.
{\it Bottom} : The ratio of CMB angular power spectra for various values 
of $\Omega_\nu$ to the one for $\Omega_\nu = 0$.  The other parameters 
are fixed at 
$\Omega_{\rm c} = 0.26$, 
$\Omega_{\rm b}=0.04$, $h=0.7$, while keeping a flat universe
$\Omega_{\rm b} + \Omega_{\rm c} + \Omega_{\nu} + \Omega_{\Lambda}  =1.0$.
The lines have the same meaning as in the figure above. 
}
\label{fig:fig5}
\end{figure}

Neutrino masses also give rise to effects in the CMB power spectrum.
If their masses are smaller than the temperature at recombination
their effect is very similar to that of massless
neutrinos \cite{crotty1}.  For slightly larger masses, there is an
enhancement of the acoustic peaks with respect to the massless case,
as shown in figures \ref{fig:fig4} and \ref{fig:fig5}.  
In figure \ref{fig:fig5} we show two cases, one (top) where 
we fix $\Omega_{\rm m}= 
\Omega_{\rm b} + \Omega_{\rm c} + \Omega_{\nu}  =0.3$, 
$\Omega_{\rm b}=0.04$, $\Omega_{\Lambda} = 0.7$ $h=0.7$.  
At the second (bottom) 
we fix 
$\Omega_{\rm c} = 0.26$, 
$\Omega_{\rm b}=0.04$, $h=0.7$, while keeping a flat universe
$\Omega_{\rm b} + \Omega_{\rm c} + \Omega_{\nu} + \Omega_{\Lambda}  =1.0$.
the purpose of this comparison is to check if the change caused by 
adding massive neutrinos
is not just due to the decrease in the amount of cold dark matter.
Indeed we see that qualitatively
the effect is the same in both cases and even 
somewhat larger when the amount of cold dark matter is held fixed.
Note that  this case the position of the
peaks will change slightly with $\Omega_{\nu}$  since varying it now also
implies changing $\Omega_\Lambda$, and hence the angular diameter distance
to the last scattering surface.  This effect is not obvious in the
figure, since we plot the ratios of the angular power spectra.

From figure
\ref{fig:fig4} it looks like the WMAP data alone should be sufficient
to provide an upper limit on the neutrino masses.  For example in the case 
of fixed
total $\Omega_{\rm m} = 0.3$ 
we find that $\Delta \chi ^ 2$ for the model with $m_\nu=0.3$ 
compared with the
reference model with $m_\nu = 0$ is larger than 40. However, note that
all other parameters have been fixed in figures \ref{fig:fig4} and
\ref{fig:fig5}, and that there are severe degeneracies between $m_\nu$
and other parameters like $n$ and $\Omega_{\rm b}h^ 2$.  
There are severe degeneracies between $m_\nu$ and other parameters
like $n$ and $\Omega_{\rm b}h^ 2$.  The full analysis of the WMAP data
alone in \cite{tegmarksdss} gave no upper limit on $m_\nu$.  On the
other hand \cite{fukugita04} have claimed an upper limit of 2.0 eV
from CMB alone, but emphasizing that one cannot improve the limit
below 1.5 eV.
The differences in the conclusions from various studies in the
literature on neutrino masses from CMB alone might be due to the
assumed priors over other cosmological parameters.


Analytic considerations in \cite{fukugita04} provide insight into the 
effect of the neutrinos on the CMB.
A key point is to consider the redshift when neutrino mass becomes 
non-relativistic, $1+z_{nr} = 6.24 \times 10^{4}  \Omega_{\nu} h^2$.
For a recombination redshift $z_{rec} = 1088$ this means that 
neutrinos became non-relativistic before recombination
$z_{nr} > z_{rec}$ if $ \Omega_{\nu} h^2 > 0.017$
(i.e.a total neutrino mass $m_{\nu, {\rm tot}} > 1.6$ eV).
In other words, 
if they are heavier than the above value, they act as matter and if 
if neutrinos are very light they nearly behave like radiation. 
Indeed as shown in \cite{fukugita04}
the dependence of the position of the first peak and the normalized 
height of the first peak  on  $ \Omega_{\nu} h^2$
has a turning point at $ \Omega_{\nu} h^2 \approx  0.017$.
This value also affects CMB anisotropy via the 
modification of the integrated Sachs-Wolfe
effect due to the massive neutrinos.

Future CMB missions like Planck
\footnote{http://www.esa.int/science/planck} will provide
high-resolution maps of the CMB temperature and polarization
anisotropies.  Gravitational lensing of these maps causes distortions,
and Kaplinghat, Knox \& Song \cite{kaplinghat} have shown that this
effect can be used to obtain very stringent limits on neutrino masses
from the CMB alone.  For Planck, they predict a sensitivity down to
$0.15\;{\rm eV}$, whereas a future experiment with higher resolution
and sensitivity can possibly reach the lower bound $\sim 0.06\;{\rm
eV}$ set by the neutrino oscillation experiments.

\section{Recent cosmological neutrino mass limits}

In an important paper 
Hu, Eisenstein \& Tegmark \cite{hutegmark} showed that one could 
obtain useful upper limits on neutrino masses from a galaxy redshift 
survey of the size and quality of the Sloan Digital Sky Survey (SDSS).  
Based on a Fisher matrix analysis, their prediction was that the SDSS 
should be able to obtain a $2\sigma$ detection of $N$ nearly 
degenerate massive neutrino species with mass 
\begin{equation}
m_\nu \geq 0.65\left(\frac{\Omega_{\rm m}h^2}{0.1N}\right)^{0.8}\;{\rm eV},  
\label{eq:hulim}
\end{equation}
which for $N=3$, $\Omega_{\rm m}h^2 = 0.135$ predicts that a 
95 \%  confidence upper limit $m_{\nu,\rm tot} \leq 1.03\;{\rm eV}$ 
should be obtainable.  

As far as we are aware of, the first cosmological neutrino mass limit after 
the detection of neutrino oscillations was derived by Croft, Hu \& Dav\'{e} 
\cite{croft}.  Their main piece of data on large-scale structure was 
the matter power spectrum derived from the Lyman-$\alpha$ forest.  
Combining with two other measurements of the amplitude of matter fluctuations, 
the COBE normalization and $\sigma_8$, they obtained a 95\% confidence 
upper limit of $m_{\nu,\rm tot} < 16.5\;{\rm eV}$, see table \ref{tab:tab1}. 
Shortly thereafter, Fukugita, Liu \& Sugiyama \cite{fukugita} derived a 
stronger upper limit of $2.7\;{\rm eV}$ by combining the 
COBE normalization of the matter power spectrum with constraints on 
$\sigma_8$ from cluster abundances,  
using strong priors $\Omega_{\rm m} < 0.4$, $h < 0.8$, and $n=1.0$.  
The first limit derived from a combined analysis of CMB and 
large-scale structure data was that of Wang, Tegmark \& Zaldarriaga 
\cite{wang1}.  They included the power spectrum of galaxies derived 
from the PSCz survey \cite{pscz} 
and obtained an upper limit of 4.2 eV.  
Going down table \ref{tab:tab1} one notes a marked improvement 
in the constraints after the 2dFGRS power spectrum became available.  
After WMAP, there is a further tendency towards stronger 
upper limits, reflecting the dual role of the CMB and large-scale 
structure in constraining neutrino masses: the matter power spectrum 
is most sensitive to the ratio $\Omega_{\nu}/\Omega_{\rm m}$, 
but one needs good constraints 
on the other relevant cosmological parameters to break degeneracies 
in order to obtain low upper mass  limits.  The limit will depend on the 
datasets and priors used in the analysis, but it seems like we 
are now converging to the precision envisaged in \cite{hutegmark}. 
In \cite{seljakbias},  galaxy-galaxy lensing was utilized  
to extract information about the linear bias parameter in the 
SDSS, making a direct association between the galaxy and matter 
power spectra, and hence resulting in a stronger constraint on the neutrino 
mass than would have been possible using just the shape of the 
galaxy power spectrum.   Finally, in \cite{seljaklyalpha} Lyman-$\alpha$ 
forest constraints were added to the data sets in \cite{seljakbias}, resulting 
in the very strong limit $m_{\nu,\rm tot} < 0.42\;{\rm eV}$ at 
95 \% confidence.
\begin{table}
\begin{center}
\begin{tabular}{l|l|l|l|l} 
\hline 
Reference & CMB & LSS  & Other& $m_{\nu,\rm tot}$ \\ 
          &     &      & data   &  upper limit          \\ \hline \hline
\cite{croft}    & ---   & Ly$\alpha$ & COBE norm., & 16.5 eV \\ 
                &       &            &$h=0.72\pm 0.08$,  &  \\
                &       &            & $\sigma_8=0.56\Omega_{\rm m}^{0.47}$ & \\
\cite{fukugita} & ---   & $\sigma_8$ &$\Omega_{\rm m}<0.4$, & 2.7 eV \\ 
                &       &            & $\Omega_{\rm b}h^2=0.015$,   \\
                &       &          & $h<0.8$, $n=1.0$&  \\ 
\cite{wang1} & pre-WMAP & PSCz, Ly$\alpha$ & --- & 4.2 eV \\ 
\cite{elgar}  & None & 2dFGRS & BBN, SNIa, & 2.2 eV \\ 
              &      &        &HST, $n=1.0\pm 0.1$ &  \\ 
\cite{steen1}      & pre-WMAP & 2dFGRS & --- & 2.5 eV  \\ 
\cite{lewis}  &pre-WMAP & 2dFGRS & SNIa, BBN &  0.9 eV \\ 
\cite{wmap}   & WMAP+CBI+ACBAR & 2dFGRS & Ly$\alpha$ &  0.71 eV \\ 
\cite{steen3}     & WMAP+Wang comp. & 2dFGRS & HST, SNIa & 1.01 eV \\ 
\cite{allen}    & WMAP+CBI+ACBAR  & 2dFGRS & X-ray & $0.56^{+0.30}_{-0.26}
\;{\rm eV}$ \\ 
\cite{tegmarksdss}   & WMAP            & SDSS   &  ---     & 1.7 eV \\ 
\cite{barger}    & WMAP            & 2dFGRS+SDSS & --- & 0.75 eV  \\ 
\cite{crotty1}    & WMAP+ACBAR      & 2dFGRS+SDSS &---  & 1.0 eV  \\
\cite{seljakbias} & WMAP            & SDSS       & bias & 0.54 eV \\ 
\cite{fukugita} & WMAP alone            & ---       & --- & 2.0 eV \\
\cite{seljaklyalpha} &WMAP       & SDSS + Ly$\alpha$& --- & 0.42 eV \\   
\hline \hline  
\end{tabular}
\label{tab:tab1}
\end{center}
\end{table} 

\section{Other cosmological probes of neutrino masses}

\subsection{The clustering amplitude}
\begin{figure}
\begin{center}
\includegraphics[width=84mm]{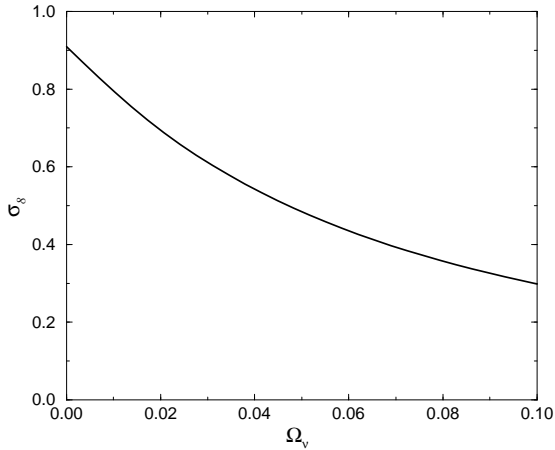}
\end{center}
\caption{The clustering amplitude $\sigma_8$ as a function of 
$\Omega_\nu$ for models with amplitude fitted to the WMAP data.
Other parameters are fixed at `concordance' values.
}
\label{fig:fig6}
\end{figure}

Direct probes of the 
total matter distribution avoid the issue of bias and are therefore 
ideally suited for providing limits on the neutrino masses. 
Several ideas for how this can be done exist.  In \cite{fukugita} 
the normalization 
of the matter power spectrum on large scales derived from COBE 
was combined with constraints on $\sigma_8$ from cluster abundances 
and a constraint $m_{\nu,\rm tot} < 2.7\;{\rm eV}$ obtained, although 
with a fairly restricted parameter space.  However, $\sigma_8$ is 
probably one of the most debated numbers in cosmology at the moment 
\cite{wang},  
and a better understanding of systematic uncertainties connected with 
the various methods for extracting it from observations is needed 
before this method can provide useful constraints.  The potential 
of this method to push the value of the mass limit down also depends 
on the actual value of $\sigma_{\rm 8}$: the higher $\sigma_8$ turns 
out to be, the less room there will be for massive neutrinos.  
As an illustration we show in figure \ref{fig:fig6} the value of 
$\sigma_8$ as a function of varying $\Omega_\nu$ with the 
remaining cosmological parameters fixed at their `concordance' values.  
For a given value of $m_\nu$, one fits the corresponding CMB power spectrum 
to the data.  This in turn leads to a best-fit amplitude and a prediction 
for $\sigma_8$ for the given value of $m_\nu$.  If one then has an 
independent measurement of $\sigma_8$, one can infer the value of 
$m_\nu$.   
In figure \ref{fig:fig6} the amplitude of the power spectrum has 
been fixed by fitting to the WMAP data.  
The claimed detection of a non-zero neutrino mass in \cite{allen}  
can be seen to be due to the use of the cluster X-ray 
luminosity function to constrain $\sigma_8$, giving $\sigma_8=0.69
\pm 0.04$ for $\Omega_{\rm m}=0.3$ \cite{allen2}.  If a value 
of $\sigma_8$ at the higher end of the results reported in the 
literature is used instead, e.g. $\sigma_8=0.9$ for $\Omega_{\rm m}=0.3$ 
from \cite{nbahcall}, one gets a very tight upper limit on 
$m_\nu$, but no detection of $m_\nu > 0$.  It is clearly important 
that systematic issues related to the various methods of obtaining 
$\sigma_8$ are settled.  The evolution of cluster abundance with 
redshift may  provide  
further constraints on neutrino masses \cite{lukash}.

\subsection{The Lyman-$\alpha$ forest} 

Simulations support the picture that the matter density field 
is locally related to the optical depth for absorption of light emitted by quasars 
by the Lyman-$\alpha$ clouds.
Hence the Lyman-$\alpha$ 
forest provides constraints on the matter power spectrum 
on scales of $k \sim 1 ~h\,{\rm Mpc}^{-1}$,
where the effect of massive neutrinos is most visible.
It was used in \cite{croft} to derive  
a limit 
$m_{\nu,\rm tot} < 16.5\;{\rm eV}$.  
However, it is non-trivial to use the information contained in the 
Lyman-$\alpha$ forest data in cosmological parameter estimation
\cite{seljak2,viel1,viel2}.  
Intriguingly, the upper limit on $m_{\nu, \rm tot}$ quoted in 
\cite{wmap} with the Lyman-$\alpha$ forest data included is actually 
{\it weaker} than the one without.    
The strongest limit to date is, as 
mentioned in the previous subsection, the one derived by Seljak et al. 
\cite{seljaklyalpha}  
by combining the power spectrum from the Lyman-$\alpha$ forest with 
SDSS data on galaxy clustering and galaxy-galaxy lensing and the 
WMAP CMB power spectrum. Such upper limit     
$m_{\nu, \rm tot} < 0.42$ already gives a non-degenerate mass hierarchy.

\subsection{Peculiar velocities}

The rms bulk flow is predicted in linear theory 
as: 
\begin{equation}
\langle v^2(R_*) \rangle =     
(2 \pi^2)^{-1}\; H_0^2 \; f^2(\Omega_{\rm m}, \Omega_\Lambda)  
\int dk P(k) W_G^2(kR_*) 
\end{equation} 
where $W_G(kR_*)$ is a window function, 
e.g. $W(kR_*) = \exp(-k^2 R_*^2/2)$  for a Gaussian sphere
of radius $R_*$. 
The  perturbations growth factor is 
$f(\Omega_{\rm m}, \Omega_\Lambda) \approx \Omega_{\rm m}^{0.6}$, hence
the peculiar velocity field is insensitive to the cosmological 
constant or dark energy (e.g. \cite{olahav2,steinhardt}).
If we parameterise the power spectrum as  $P(k) \equiv A k^n T^2(k) $ 
we can then consider two model universes with the same early universe
primordial power spectrum $A k^n$ but different transfer functions
$T(k)$ according to the contribution of the neutrinos.
We define to total mass density parameter as
$\Omega_{\rm m} = \Omega_{\rm c} + \Omega_{\rm b} + \Omega_{\nu}$
due to the contributions of cold dark matter and neutrinos.

\begin{figure}
\begin{center}
\includegraphics[width=84mm]{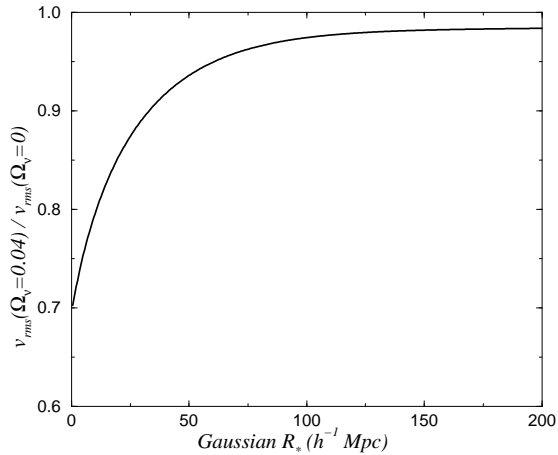}
\end{center}
\caption{
The ratio of the bulk flows in Gaussian spheres for two universes with the same
primordial power spectra, same $\Omega_{\rm m} =0.3; 
\Omega_{\rm b}=0.04; h=0.7; n=1$ 
but different $\Omega_{\nu} = 0 $ and 0.04.
}
\label{fig:pecvel_ratio}
\end{figure}

Figure \ref{fig:pecvel_ratio}
shows the ratio of the bulk flows in two universes with the same
primordial power spectra, same $\Omega_{\rm m} =0.3; 
\Omega_{\rm b}=0.04; h=0.7; n=1$ 
but different $\Omega_{\nu} = 0 $ and 0.04.
we can see the suppression of $\sim 20 \%$ of velocities
on radii $R_* < 50 h^{-1}$ Mpc due to the massive neutrinos.

\begin{figure}
\begin{center}
\includegraphics[width=84mm]{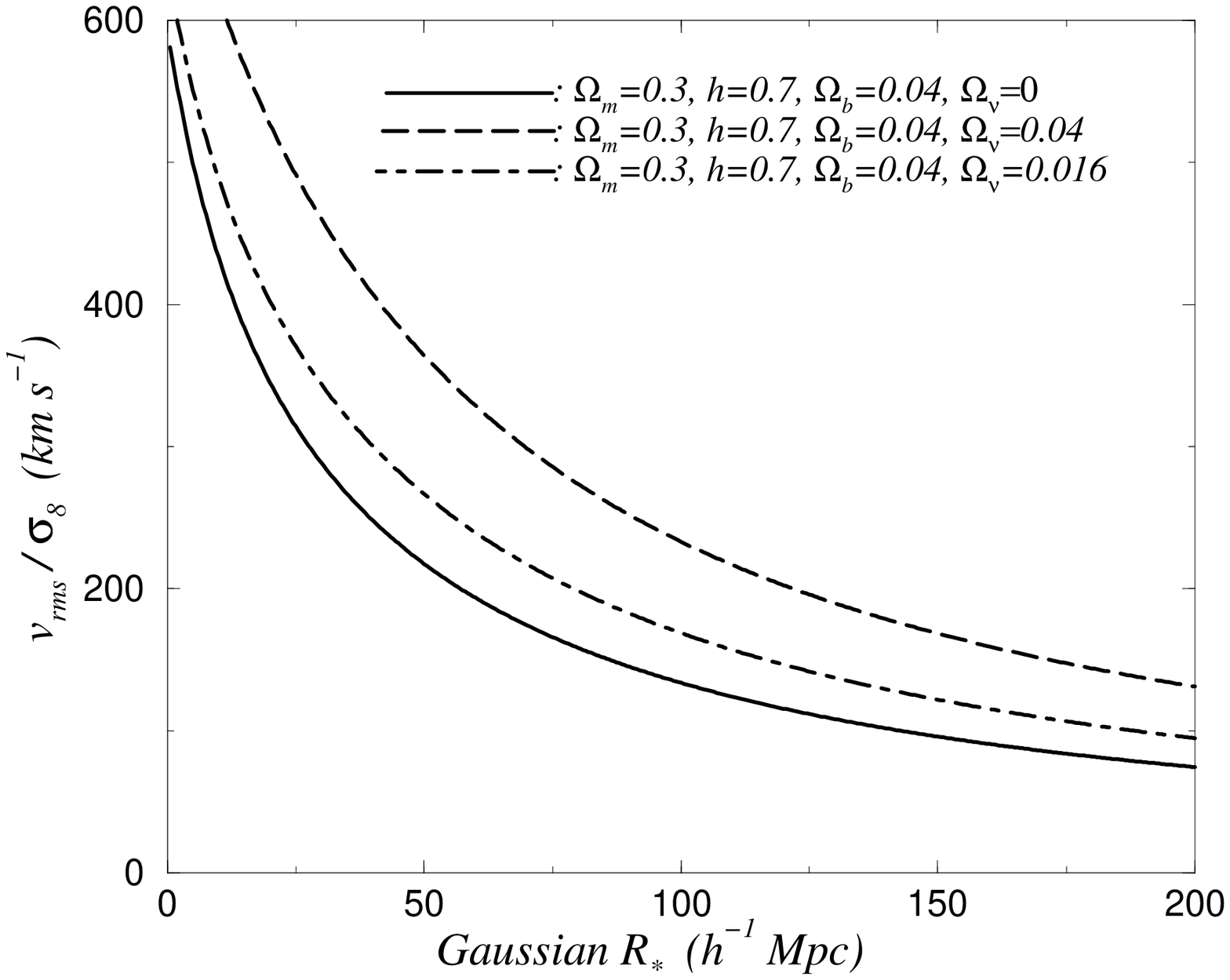}
\includegraphics[width=84mm]{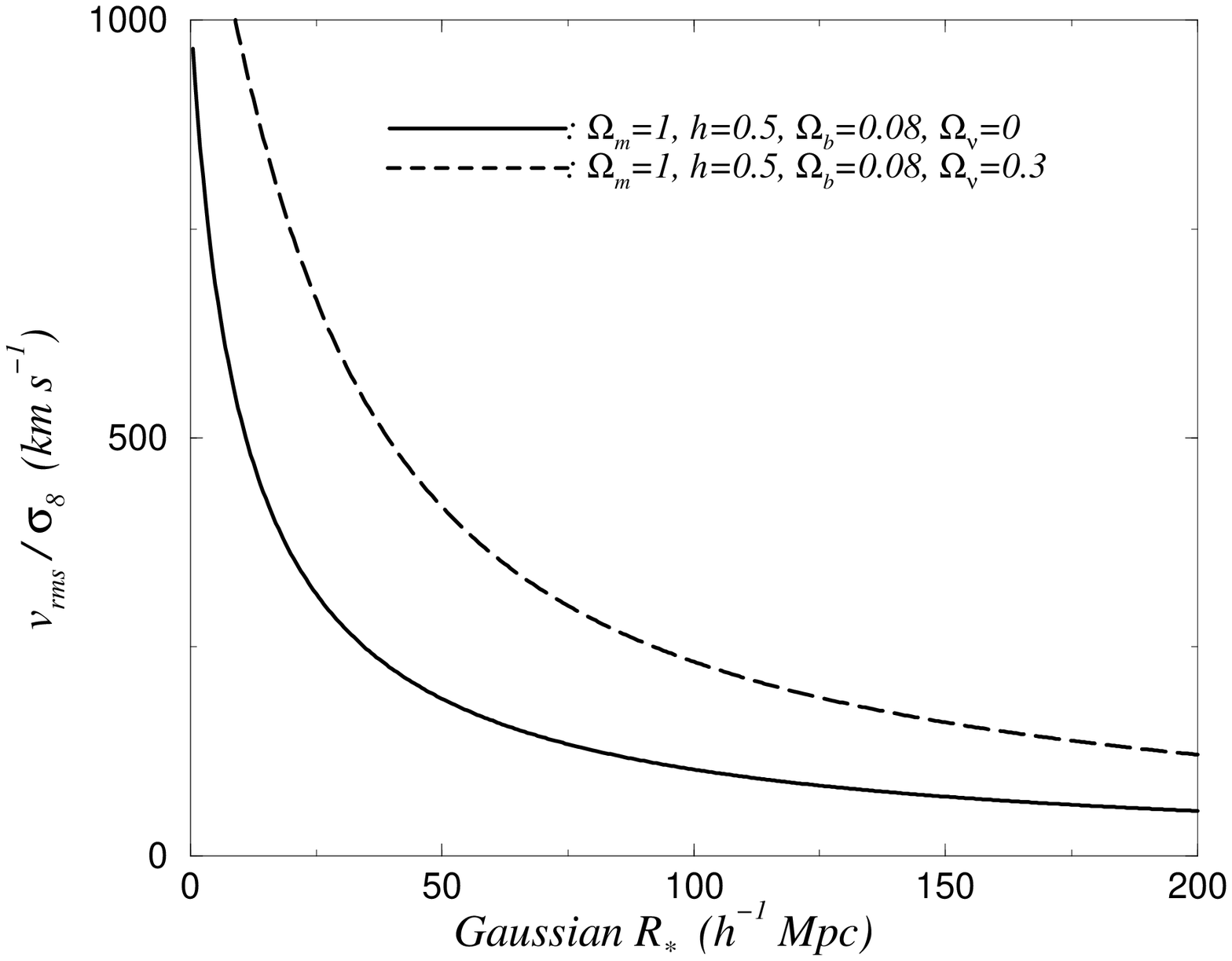}
\end{center}
\caption{The effect of massive neutrinos on peculiar velocities in 
$\Lambda{\rm CDM}$ models (top panel) and MDM models (bottom panel).}
\label{fig:fig10}
\end{figure}

If the  power spectrum normalization  $\sigma_8$  is known from another
independent measurement, 
then the amplitude $A$ itself is inversely proportional to 
an integral over the power spectrum as
\begin{equation}
\sigma^2_8 =     (2 \pi^2)^{-1}A 
\int dk  k^{2} k^n T^2(k)  W_{TH}^2(8k) 
\end{equation}
where $W_{TH}$ is an $8\;h^{-1}$ Mpc radius spherical top-hat window
function.  One can then make predictions for the bulk flows
$v_{rms}/\sigma_8$ in different models, as shown in Figure
\ref{fig:fig10} In part the dramatic differences for models with and
without massive neutrinos are due to the $\sigma_8$ normalization
procedure.  While it is encouraging that the effect of massive
neutrinos is significant for small scales, where peculiar velocities
can be measured more accurately to nearby galaxies, we should be aware
of complications due to non-linear effects and systematic errors.

\subsection{Gravitational Lensing}
Deep and wide weak lensing 
surveys will in the future make it possible to perform weak lensing tomography 
of the matter density field \cite{hulens1,hulens2}.  By binning 
the galaxies in a deep and wide survey in redshift, one can probe the 
evolution of the gravitational potential.  However, because massive 
neutrinos and dark energy have similar effects on this evolution, 
complementary information is required in order to break this degeneracy.  
Several studies of the potential of lensing tomography to constrain 
cosmological parameters, in particular dark energy and neutrino masses, 
have been carried out, see e.g. \cite{knox} for an overview.  
Even when taking the uncertainties in the properties of dark energy 
into account, the combination of weak lensing tomography and 
high-precision CMB experiments may reach sensitivities down to the 
lower bound of 0.06 eV on the sum of the neutrino masses 
set by the current oscillation data \cite{knox}.

\section{Discussion}

The dramatic increase in amount and quality of CMB and large-scale 
structure data we have seen in cosmology in the last few years 
have made it possible to derive fairly stringent limits on the 
neutrino mass scale.  Even though the CMB is less sensitive to neutrino 
masses than large-scale structure data, it plays a crucial role in 
breaking parameter degeneracies. With the WMAP and SDSS data, the upper limit 
has been pushed down to $\sim 1\;{\rm eV}$ for the total mass,  
assuming three massive neutrino species. 
Results from Lyman-$\alpha$ clouds combined with the WMAP and SDSS
galaxy clustering data provide a tighter upper limit of $\sim 0.4\;{\rm eV}$.

One point to bear in mind is that all these limits assume the 
`concordance' $\Lambda{\rm CDM}$ model with adiabatic, scale-free 
primordial fluctuations.  While the wealth of cosmological data 
strongly indicate that this is the consistent  basic picture, one 
should keep in mind that cosmological neutrino mass limits 
are model-dependent, and that there might still be surprises.
Foe example,
as the suppression of the power spectrum depends on the ratio
$\Omega_{\nu}/\Omega_{\rm m}$, we found in \cite{elgar} that the
out-of-fashion Mixed Dark Matter (MDM) model, with $\Omega_{\nu}=0.2$,
$\Omega_{\rm m}=1 $ and no cosmological constant, fits the 2dFGRS power
spectrum well, but only for a Hubble constant $H_0 < 50\;{\rm km}\,{\rm
s}^{-1}\,{\rm Mpc}^{-1}$. 
A similar conclusion was reached in 
\cite{blanchard}, and they also found that the CMB power spectrum could be
fitted well by the same MDM model if one allows features in the
primordial power spectrum.  
Another consequence of
this is that excluding low values of the Hubble constant, e.g. with
the HST Key Project, is important in order to get a strong upper limit
on the neutrino masses.

If the future observations live up to their promise, 
the prospects for pushing the cosmological neutrino mass limit 
down towards $0.1\;{\rm eV}$ are good.  Then, 
as pointed out in 
\cite{lesgourgueslatest}, one may even  
start to see effects of the different mass hierarchies (normal or 
inverted), and thus one should take this into account when 
calculating CMB and matter power spectra.  For example, with a 
non-degenerate mass hierarchy one will get more than one free-streaming 
scale, and this will leave an imprint on the matter power spectrum.  
The coming years will see 
further comparison between the effective neutrino mass in Tritium beta decay, 
the effective Majorana neutrino mass in neutrinoless double beta decay and the 
sum of the neutrino masses from cosmology \cite{fogli04}, \cite{zuber}.
It would be a great triumph for cosmology if the neutrino mass 
hierarchy were finally revealed by the distribution of large-scale 
structures in the Universe.

\ack  
We thank S. Bridle and M. Fukugita for valuable discussions.  
{\O}E acknowledges support from the Research Council of Norway
through grant number 159637/V30 and
OL thanks PPARC for a Senior Research Fellowship.

\end{document}